\documentclass{article}[draft]
\usepackage{setspace}
\usepackage{graphicx} 
\usepackage{authblk}
\usepackage{mwe} 
\usepackage{blindtext} 
\usepackage{placeins}
\usepackage{siunitx} 
\usepackage{mathtools} 
\usepackage{chemformula} 
\usepackage[utf8]{inputenc}

\usepackage[style=phys,sorting=none]{biblatex} 
\usepackage{amsmath,amssymb}
\usepackage{bm}
\usepackage{listings} 
\usepackage{graphicx}
\usepackage{lmodern}
\usepackage[textwidth=16cm,textheight=23cm]{geometry}
\usepackage{braket}

\usepackage{url}
\usepackage[colorlinks=true, pdfstartview=FitV, linkcolor=blue, citecolor=blue, urlcolor=blue]{hyperref}

\usepackage[font=small, labelfont=bf]{caption}
\usepackage{subfig} 
\usepackage{float}

\usepackage{booktabs}   
\usepackage{longtable}
\usepackage{array}
\usepackage{threeparttablex}

\newcolumntype{x}[1]{%
>{\centering\hspace{0pt}}p{#1}}%

\usepackage{enumitem}
\usepackage{xr}
\author[1]{Dániel Nagy}
\author[2]{Peter Reinholdt}
\author[1]{Phillip W. K. Jensen}
\author[2]{Erik Rosendahl Kjellgren}
\author[3]{Karl Michael Ziems}
\author[4]{Aaron Fitzpatrick}
\author[4,5]{Stefan Knecht}
\author[2]{Jacob Kongsted}
\author[3]{Sonia Coriani}
\author[1]{Stephan P. A. Sauer}
\affil[1]{\textit{Department of Chemistry, University of Copenhagen, DK-2100 Copenhagen {\O}, Denmark}} 
\affil[2]{\textit{Department of Physics, Chemistry and Pharmacy, University of Southern Denmark, DK-5230 Odense M, Denmark}}
\affil[3]{\textit{DTU Chemistry, Technical University of Denmark, DK-2800, Kongens Lyngby, Denmark}}
\affil[4]{\textit{Algorithmiq Ltd, Kanavakatu 3C, FI-00160 Helsinki, Finland}}
\affil[5]{\textit{ETH Z{\"u}rich, Department of Chemistry and Applied Life Sciences, CH-8093 Z{\"u}rich, Switzerland}}

\date{}
\title{Electric Field Gradient 
Calculations for Ice VIII and IX using Polarizable Embedding: A Comparative Study on Classical Computers and Quantum Simulators}

\begin{document}
\maketitle
\begin{abstract}
We test the performance of the Polarizable Embedding Variational Quantum Eigensolver Self-Consistent-Field (PE-VQE-SCF) model for computing electric field gradients with comparisons to conventional complete active space self-consistent-field (CASSCF) calculations and experimental results. We compute quadrupole coupling constants for ice VIII and ice IX.  We find that the inclusion of the environment is crucial for obtaining results that match the experimental data. The calculations for ice VIII are within the experimental uncertainty for both CASSCF and VQE-SCF for oxygen and lie close to the experimental value for ice IX as well.
With the VQE-SCF, which is based on an Adaptive Derivative-Assembled Problem-Tailored (ADAPT) ansatz, we find that the inclusion of the environment and the size of the different basis sets do not directly affect the gate counts. However, by including an explicit environment, the wavefunction and, therefore, the optimization problem 
becomes more complicated, which usually results in the need to include more operators from the operator pool, thereby increasing the depth of the circuit.
\end{abstract}


\section{Introduction}


In the realm of theoretical chemistry, limitations in the computational resources that classical computers can provide often become a major bottleneck, that inhibits accurate calculations, even for medium-sized molecules and systems. Quantum chemists have made huge efforts to develop different approximations and more effective algorithms to reduce the computational cost while gaining accurate results. Therefore, it is not surprising that the advance of quantum computers drew a lot of attention in life sciences and quantum chemistry. 
A number of different comprehensive reviews discuss the current state of development of theories, methods, and algorithms that focus on performing quantum mechanical calculations on near-term noisy intermediate-scale quantum (NISQ) devices and on far-future fault-tolerant ones \cite{Cao2019, Bauer2020, Elfving2020, McArdle2020b, Motta2022, Aulicino2022, Bharti2022, Liu2022}.
\\[2mm]
To circumvent the noisy nature of the currently available quantum devices, hybrid quantum-classical algorithms have been developed \cite{McClean2016}, where only certain parts of the calculation are executed on a quantum computer. The remaining parts are carried out on classical CPU- or GPU-based computers. The tasks that are executed on quantum computers are mostly expectation-value measurements of the Hamiltonian or other quantum mechanical operators. Within the field of chemistry applications, the Variational Quantum Eigensolver (VQE) \cite{Cao2019,Yung2014,Romero2019, Wang2019, Fedorov2022b} algorithm is one of the most popular choices. \\[2mm]
Chemical experiments are rarely done on isolated molecules, so it can be crucial to include environmental effects to obtain meaningful and reliable results that reproduce experiments. 
One of the most successful and feasible approaches to model such effects describes the ``active'' part of a system with a quantum mechanical method, while the environment is described in a more approximate manner.
For classical computers, many such hybrid methods have been developed and implemented, ranging from simple continuum solvent models like the Polarizable Continuum Model (PCM)~\cite{Mennucci2012} to the more accurate treatments like the Polarizable Embedding (PE) model~\cite{olsen2010, olsen2011}. 
\\[2mm]
The electric field gradient (EFG)  is very sensitive to the charge distribution of the nucleus environment due to its $r^{-3}$ dependency~\cite{SauerMolelem2011}, and therefore it is a challenging property to calculate accurately even on a classical computer. The EFG is needed to calculate the nuclear coupling interaction (NQI) that is important in the interpretation of different spectroscopic measurements such as perturbed angular correlation (PAC) or M{\"o}ssbauer spectroscopy~\cite{spas051, C2CP23080, C2CP42291C, Haas_2017, D2CP05574K}.\\[2mm]
In this paper, we will use an implementation of the PE model for quantum computers \cite{Kjellgren2023} in combination with the Adaptive Derivative-Assembled Problem-Tailored Ansatz Variational Quantum Eigensolver self-consistent field approach (ADAPT-VQE-SCF)~\cite{Fitzpatrick2022} and test its performance on calculating the EFGs of ice VIII and ice IX. 
We aim to reproduce the experimental results with conventional CASSCF and the ADAPT-VQE-SCF approach on a quantum simulator for ice VIII and ice IX.

\section{Theory}
\label{Sec:Theory}
In the following, we summarize the key ingredients of the theoretical background for this study. The summary is divided into three parts: first, we define the electric field gradient; second, we introduce the polarizable embedding model; and third, we describe the VQE-SCF scheme with the inclusion of the PE model.

\subsection{The electric field gradient at the nucleus}
The EFG can not be measured directly in experiments. To compare our calculations to experimental results, we instead compare the calculated and measured nuclear quadrupole interactions (NQI). In the following, we will briefly describe EFG and direct the reader to Ref.~\cite{SauerMolelem2011} (Chapter 4) for a more detailed introduction to EFG.
\\[2mm]
The NQI is defined by the coupling between the nuclear quadrupole moment and the EFG
\begin{equation} \label{eq2.4}
    \chi = \frac{Q}{2\pi}\varepsilon_{zz}~,
\end{equation}
where $\chi$ is the NQI, $Q$ is the electric quadrupole moment of the nucleus, and $\varepsilon_{ii}$ with $i=x,y,z$ are the eigenvalues of the EFG tensor, ordered such that $|\varepsilon_{zz}|\geq|\varepsilon_{yy}|\geq|\varepsilon_{xx}|$. We note that $\varepsilon_{zz}=-(\varepsilon_{xx}+\varepsilon_{yy})$, since the EFG tensor is traceless. 
From the three eigenvalues, one can define an asymmetry parameter:

\begin{equation}
    \eta_K = \frac{\varepsilon_{xx}-\varepsilon_{yy}}{\varepsilon_{zz}}~,
\end{equation}
at the position of nucleus $K$. The electric field gradient is the second derivative of the electrostatic potential
\begin{equation}
    \varepsilon_{\alpha\beta}(\bm{\mathrm{{R_0}}}) = -\frac{\partial^2\phi^\varepsilon(\bm{\mathrm{{r}}})}{\partial r_{\alpha}\partial r_{\beta}}\Bigm\lvert_{\bm{\mathrm{r}}=\bm{\mathrm{{R_0}}}}~,
\end{equation}
where $\varepsilon_{\alpha\beta}(\bm{\mathrm{{R_0}}})$ is the electric field gradient tensor evaluated at $\bm{\mathrm{{R_0}}}$, $\phi^\varepsilon(\bm{\mathrm{{r}}})$ is the electrostatic potential, $\alpha$ and $\beta$ are Cartesian coordinates.
The quantum mechanical expression for the electric field gradient tensor at nucleus $K$ is
\begin{equation}
\begin{gathered}
     \varepsilon_{\alpha\beta}(\bm{\mathrm{{R}}}_K) = \left\langle\Psi_0 \middle|  \sum_i^N\left[3\frac{(\bm{\mathrm{{r}}}_{i,\alpha}-\bm{\mathrm{{R}}}_{K,\alpha})(\bm{\mathrm{{r}}}_{i,\beta}-\bm{\mathrm{{R}}}_{K,\beta})}{|\bm{\mathrm{{r}}}_i-\bm{\mathrm{{R}}}_K|^5} - \frac{\delta_{\alpha\beta}}{|\bm{\mathrm{{r}}}_i-\bm{\mathrm{{R}}}_K|^3}\right]\middle| \Psi_0\right\rangle \\
      - \sum_{L\neq K}Z_L\left[3\frac{(\bm{\mathrm{R}}_{L,\alpha}-\bm{\mathrm{R}}_{K,\alpha})(\bm{\mathrm{R}}_{L,\beta}-\bm{\mathrm{R}}_{K,\beta})}{|\bm{\mathrm{R}}_L-\bm{\mathrm{R}}_K|^5} - \frac{\delta_{\alpha\beta}}{|\bm{\mathrm{R}}_L-\bm{\mathrm{R}}_K|^3}\right] \label{eq:EFG_tensor}
\end{gathered},
\end{equation}
where $\bm{\mathrm{{r}}}_i$ are the positions of the electrons and $\bm{\mathrm{{R}}}_K$ are the positions of the nuclei, $\delta_{\alpha\beta}$ is the Kronecker delta, and $\ket{\Psi_0}$ is the ground state wavefunction. The first term describes the EFG due to the electrons, while the second describes the nuclear part of the EFG. \\
In second quantization, one can obtain the electric field gradient as the one-electron reduced density matrix (1-RDM) of the system contracted with the one-electron integrals of the property, plus the nuclear contribution:

\begin{equation}
    \label{efg_secondquant}
    \varepsilon_{\alpha\beta}(\bm{\mathrm{R}}_K) = \textbf{f}_{\alpha\beta}(\bm{\mathrm{R}}_K)\bm{\mathrm{D}} - \sum_{L\neq K}Z_L\left[3\frac{(\bm{\mathrm{R}}_{L,\alpha}-\bm{\mathrm{R}}_{K,\alpha})(\bm{\mathrm{R}}_{L,\beta}-\bm{\mathrm{R}}_{K,\beta})}{|\bm{\mathrm{R}}_L-\bm{\mathrm{R}}_K|^5} - \frac{\delta_{\alpha\beta}}{|\bm{\mathrm{R}}_L-\bm{\mathrm{R}}_K|^3}\right]~,
\end{equation}

where $\textbf{f}_{\alpha\beta}(\bm{\mathrm{R}}_K)$ is the matrix of one-electron integrals of the property ($\alpha\beta$ component thereof) evaluated at the position of nuclei $K$ and $\bm{\mathrm{D}}$ is the reduced one-electron density matrix (1-RDM).
The EFG has been calculated from the optimized wavefunctions according to 
equation~\ref{efg_secondquant}.\\


Since the electric field gradient has a $r^{-3}$ dependence, it is a very local molecular property and highly sensitive to the surrounding environment. Therefore, comparing our calculations to experimental results, including the environmental contributions is crucial. 

\subsection{Polarizable Embedding}
\label{Sec:PE}
In the Polarizable Embedding~\cite{olsen2010, olsen2011} (PE) model, a molecular system is divided into two parts: a core region that is treated quantum mechanically (QM) and a surrounding environment, where the molecules are treated semi-classically (MM) through a combination of atomically distributed point multipoles and an induced dipole model. 
In the following, we will briefly overview the PE model and refer the reader to Refs. \cite{Steinmann2019,Kjellgren2023} for further details.
\\[2mm]

The total PE energy of the system can be divided into 
\begin{equation}
    E_{\mathrm{emb}} = E_{\mathrm{vac}} + E_{\mathrm{PE}}~,
\end{equation}
where $E_{\mathrm{vac}}$ stands for the energy of the core region in a vacuum, and $E_{\mathrm{PE}}$ stands for the energy that comes from the environment.
The Hamiltonian of the whole system can be split into two parts:

\begin{equation}
    \hat{H}_{\mathrm{emb}} = \hat{H}_{\mathrm{vac}} + \hat{\upsilon}_{\mathrm{PE}}~,
    \label{eq.total.Hamilton}
\end{equation}
where $\hat{H}_{\mathrm{vac}}$ represents the isolated Hamiltonian for the quantum region without the presence of the environment. The 
$\hat{\upsilon}_{\mathrm{PE}}$ operator can be written as the sum of an electrostatic operator ($\hat{\upsilon}_{\mathrm{es}}$) that describes the potentials of the permanent charge distributions of the environment, and an induction operator ($\hat{\upsilon}_{\mathrm{ind}}$)

\begin{equation}
    \hat{\upsilon}_{\mathrm{PE}} =  \hat{\upsilon}_{\mathrm{es}} +  \hat{\upsilon}_{\mathrm{ind}}~.
\end{equation}

The electrostatic operator can be written as

\begin{equation}
     \hat{\upsilon}_{\mathrm{es}} = \sum_{s=1}^N\sum_{|k| = 0}^K\frac{(-1)^{|k|}}{k!}M_\mathrm{s}^{(k)}\hat{V}_{\mathrm{s,el}}^{(k)}~,
\end{equation}
where $M_s^{(k)}$ is the $k$'th Cartesian component of the multipoles on expansion site \emph{s}, and $\hat{V}_{s,\textrm{el}}^{(k)}$ is the potential derivative operator at site \emph{s}. When truncating the multipole expansion to include up to quadrupoles, the electrostatic operator is thus

\begin{equation}
     \hat{\upsilon}_{\mathrm{es}} = \sum_{s=1}^N\left(q_s\hat{V}_{\mathrm{s,\textrm{el}}}-\sum_{\alpha}\mu_s^\alpha\hat{V}_{\mathrm{s,el}}^{\alpha}+\sum_{\alpha\beta}\Theta_s^{\alpha\beta}\hat{V}_{s,\mathrm{el}}^{\alpha\beta}\right),
\end{equation}

where $\alpha$ and $\beta$ stand for Cartesian coordinates; $q_s$, $\mu_s^\alpha$ and $\Theta_s^{\alpha\beta}$ are atomic charges, static dipole, and quadrupole moments, respectively. The 
$\hat{V}_{s,\textrm{el}}^{(k)}$ operator can be defined in second quantization as

\begin{equation}
     \hat{V}_{s,\mathrm{el}}^{(k)} = \sum_{pq}t_{pq}^{(k)}(\bm{\mathrm{R}}_s)\hat{E}_{pq}~,
\end{equation}

where $t_{pq}^{(k)}(\bm{\mathrm{R}}_s)$ are the potential derivative integrals over molecular orbitals $\phi_p(\bm{\mathrm{r}})$, where $k$ is the order of the derivative

\begin{equation}
     t_{pq}^{(k)}(\bm{\mathrm{R}}_s) = \int\phi_p^*(\bm{\mathrm{r}})\nabla^k\left( \frac{1}{|\bm{\mathrm{r}}-\bm{\mathrm{R}}_s|}\right)\phi_q(\bm{\mathrm{r}})\,\mathrm{d}\bm{\mathrm{r}}~,
\end{equation}

and $\hat{E}_{pq}=\hat{a}^{\dag}_{p\alpha}\hat{a}_{q\alpha}+\hat{a}^{\dag}_{p\beta}\hat{a}_{q\beta}$ are second quantization singlet one-electron excitation operators. $\nabla^k$ is defined as $\frac{\partial^{|k|}}{\partial x^{k_x}\partial y^{k_y}\partial z^{k_z}}$.

The induction operator expresses the effect of a polarized charge distribution and can be defined as
\begin{equation}
     \hat{\upsilon}_{\mathrm{ind}} = -\sum_{s=1}^N\bm{\mu}_s^{\mathrm{ind}}(\bm{\mathrm{F}}_{\mathrm{tot}})\hat{F}_{\mathrm{s,el}}
\end{equation}
where $\hat{F}_{s,\textrm{el}}$ is the operator that yields the electronic electric field at site $s$, and $\boldsymbol{\mu}_s^{\mathrm{ind}}(\mathbf{F}_{\mathrm{tot}})$ are the induced dipole moments at site $s$. The induced dipoles depend on the total electric field and the polarizability at polarizable site $s$. The total electric field comes from the field of the nuclei and electrons from the quantum region, the permanent multipoles, and the induced dipoles from the environment.
\\[2mm]
To build the  
$\hat{\upsilon}_{\mathrm{ind}}$ operator, we need the induced dipoles 
$\boldsymbol{\mu}_s^{\mathrm{ind}} (\bm{\mathrm{F}}_{\mathrm{tot}})$. The induced dipoles satisfy

\begin{equation}
\label{inddip}
     \bm{\mu}_s^{\mathrm{ind}}(\bm{\mathrm{F}}_{\mathrm{tot}}) = \bm{\alpha}_s\bm{\mathrm{F}}_{\mathrm{tot}}({\bm{\mathrm{R}}}_s) = \bm{\alpha}_s\left(
     \bm{\mathrm{F}}({\bm{\mathrm{R}}}_s)
     +\sum_{s^{'}\neq s} \bm{\mathrm{T}}^{(2)}_{ss^{'}}  \bm{\mathrm{\mu}}^{\mathrm{ind}}_{s^{'}}\right)~.
\end{equation}

In equation~\eqref{inddip}, $\bm{\mathrm{F}}$ is the electric field at site $s$ from the nuclei, electrons, and permanent multiple moments, but without the induced dipoles, and $\bm{\mathrm{T}}^{(2)}_{ss^{'}}$ is the dipole-dipole interaction tensor \cite{Stone2013Theory}. For the sake of clarity, we introduce a column vector containing the induced dipoles $ \bm{\mu}^{\mathrm{ind}} = (\bm{\mu}_1^{\mathrm{ind}},\bm{\mu}_2^{\mathrm{ind}}, ... ,\bm{\mu}_N^{\mathrm{ind}})^T$, and one containing the electric fields $\bm{\mathrm{F}} = (\bm{\mathrm{F}}(\bm{\mathrm{R}}_1),\bm{\mathrm{F}}(\bm{\mathrm{R}}_2), ... , \bm{\mathrm{F}}(\bm{\mathrm{R}}_N))^T$. 
We can thus rewrite 
equation~\eqref{inddip} as

\begin{equation}
     \bm{\mathrm{B}}\bm{\mu}^{\mathrm{ind}} = \bm{\mathrm{F}}~,
\end{equation}

where $\mathbf{B}$ is the ($3N\times3N$) classical response matrix

\begin{equation}
     \mathbf{B} = \begin{pmatrix}
\bm{\alpha}^{-1}_1 & -\bm{\mathrm{T}}^{(2)}_{12} & \cdots & -\bm{\mathrm{T}}^{(2)}_{1N}\\
-\bm{\mathrm{T}}^{(2)}_{21} & \bm{\alpha}^{-1}_2 & \ddots & \vdots\\
 \vdots & \ddots & \ddots & -\bm{\mathrm{T}}^{(2)}_{(N-1)N}\\
-\bm{\mathrm{T}}^{(2)}_{N1} & \cdots & -\bm{\mathrm{T}}^{(2)}_{N(N-1)} & \bm{\alpha}^{-1}_N
\end{pmatrix}~.
\end{equation}
\\[2mm]
Since the induced dipoles depend on the electric field from the electron density of the quantum region, and the electron density depends on the induced dipoles of the environment, equation (\ref{inddip}) leads to a set of coupled equations that can be solved in a dual self-consistent manner.\\

Including an environment in our calculations will change the wave function of the quantum region, which will, in turn, lead to a change in the calculated molecular property. 
The inclusion of the environment also leads to a direct contribution to the EFG since the permanent and induced charge distributions of the environment also create electric field gradients themselves.
The EFG contribution from the environment at some point $\mathbf{R}_p$ is

\begin{equation}
    \varepsilon_{\alpha\beta} (\mathbf{R}_p) = \sum_{s=1}^{N} \sum_{|k|=0}^{K}  \frac{(-1)^{|k|}}{k!} {M}_s^{(k)}{T}^{(k+2)}_{ps,\alpha\beta...}  + \frac{1}{2} \sum_{s=1}^{N} \sum_{\gamma=x,y,z} \mu^{\mathrm{ind}}_{s,\gamma} {T}^{(3)}_{ps,\alpha\beta\gamma}~,
\end{equation}

where the term due to the static multipoles is independent of the QM density, while the term due to the induced dipoles depends on QM density.

\subsection{PE-VQE-SCF}
\label{Sec:PE-VQE-SCF}
In the following, we introduce the PE-VQE-SCF formalism, developed by Kjellgren \emph{et al.}~\cite{Kjellgren2023}, which builds upon the ADAPT-VQE-SCF originally developed by Fitzpatrick \emph{et al.} \cite{Fitzpatrick2022}. 
Further information can also be found in Ref.~\cite{Kjellgren2023}.
We begin by expressing the spin-free, non-relativistic, electronic Hamiltonian in second quantization as 

\begin{equation} \label{mol.ham}
     \hat{H} = \sum_{pq}h_{pq}\hat{E}_{pq} + \frac{1}{2}\sum_{pqrs}g_{qprs}(\hat{E}_{pq}\hat{E}_{rs}-\delta_{qr}\hat{E}_{ps})
\end{equation}

where $\hat{E}_{pq}$ is again a singlet one-electron excitation operator, and $h_{pq}$ and $g_{pqrs}$ are,
respectively, the one- and two-electron integrals over the molecular orbitals $\phi_p(\textbf{r})$

\begin{equation}
     h_{pq} = \int \phi_p^*(\textbf{r})\hat{h}\phi_q(\textbf{r})\,\mathrm{d}\textbf{r}~,
\end{equation}

\begin{equation}
     g_{pqrs} = \int \phi_p^*(\mathbf{r}_1)\phi_r^*(\mathbf{r}_2)r_{12}^{-1}\phi_q(\mathbf{r}_1)\phi_s(\mathbf{r}_2)\,\mathrm{d}\mathbf{r}_1\mathrm{d}\mathbf{r}_2~.
\end{equation}
\\[1mm]
In conventional CASSCF multi-configurational self-consistent field (MCSCF) methods~\cite{Helgaker2013-xk}, the wavefunction is parameterized by orbital rotation coefficients ($\kappa_{pq}$) and configurational interaction (CI) coefficients ($\theta_i$). The parameterized wavefunction can be written as:

\begin{equation}
     |\Psi(\bm{\kappa},\bm{\theta})\rangle = U(\bm{\kappa})U(\bm{\theta})|0\rangle~,
\end{equation}

where $|0\rangle$ is the reference state. The orbital rotations are parametrized with

\begin{equation}
     U(\bm{\kappa}) = \mathrm{exp}(\hat{\kappa}),
\end{equation}

where $\hat{\kappa}$ is the anti-hermitian orbital rotation operator that ensures orthonormal MOs during optimization. It takes the form:

\begin{equation}
     \hat{\kappa} = \sum_{p>q}\kappa_{pq}\left(\hat{E}_{pq}-\hat{E}_{qp}  \right).
\end{equation}

The difference between classical MCSCF and VQE-SCF parameterization is in the description of CI coefficients. In MCSCF, the wavefunction
is expressed as a linear combination of different Slater determinants with CI coefficients acting as weights.
However, the CI expansion must be expressed as unitary qubit rotations to target a quantum computer.
In the VQE-SCF procedure, we optimize the CI coefficients on the quantum computer and the orbital rotations on the classical computer. By having the CI coefficients present only in the active space while the orbital rotations cover the remaining orbitals, we obtain a more optimal way of dividing the tasks between the classical and quantum processors, which requires fewer qubits. 

In practice, this means that we only measure the active part of the 1- and 2-RDM on a quantum computer. The whole 1-RDM therefore has to be built up by adding the inactive occupied and unoccupied parts to the matrix.

The unitary qubit rotations are parameterized as

\begin{equation}
     U(\bm{\theta}) = \prod_i \mathrm{exp}(i\theta_{i}\hat{P}_{i})
\end{equation}

where $\hat{P}_{i}$ are strings of Pauli operators. The optimization process 
occurs
by minimizing some energy expectation value

\begin{equation}
\label{energy}
     E = \mathrm{\min_{\bm{\theta},\bm{\kappa}}} \langle0|U^{\dag}(\bm{\theta})U^{\dag}(\bm{\kappa})\hat{H}U(\bm{\kappa})U(\bm{\theta}) |0\rangle~.
\end{equation}
with respect to the orbital rotation coefficients on the classical computer, and with respect to the unitary qubit rotations on a quantum computer.

When we include a PE environment, 
we actually minimize 
the free energy in solution, 

\begin{equation}
     E_{\mathrm{FE}} = \mathrm{\min_{\bm{\theta},\bm{\kappa}}} \langle0|U^{\dag}(\bm{\theta})U^{\dag}(\bm{\kappa})\hat{H}_{\mathrm{FE}}U(\bm{\kappa})U(\bm{\theta}) |0\rangle,
\end{equation}
i.e., the expectation value of the free-energy Hamiltonian in solution, 
\begin{equation}
\hat{H}_{\mathrm{FE}} = \hat{H}_{\mathrm{emb}}  -\frac{1}{2}\hat{\upsilon}_{\mathrm{ind}}~,
\end{equation}
instead of 
the expectation value in equation~\eqref{energy}
of the molecular hamiltonian in equation (\ref{mol.ham}).


\section{Computational details}
\label{Sec:CompDet}
\begin{figure}[h!]
    \centering
    \subfloat{\includegraphics[width=7.5cm]{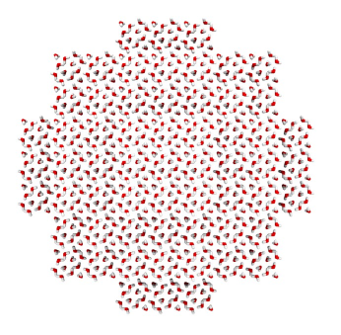}}
    \subfloat{\includegraphics[width=7.5cm]{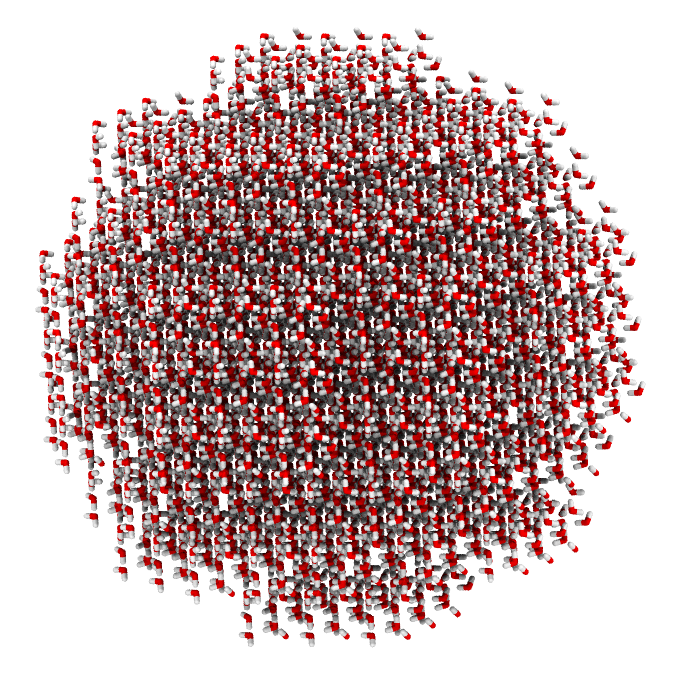}}
    \caption{The ice IX system used in the calculation. All cells included within 30 Å. Left: view along a cell vector, right: rotated view.}
    \label{fig:ice9_super_plain}
\end{figure}

The geometries utilized in all calculations are taken 
from Ref.~\cite{Santra2013}, which are experimental geometries that are relaxed with density functional theory (DFT) using the PBE0~\cite{Adamo1999} exchange-correlation functional, keeping the experimental cell dimensions. The unit cell of ice VIII contains 8 water molecules, while ice IX contains 12 water molecules. In both cases, we generated large supercells and included every unit cell within 30 {\AA} of the central one. Figure~\ref{fig:ice9_super_plain} shows an example of such a system. The ice VIII and ice IX systems contain 5832 and 8748 water molecules, respectively. 
For the central unit cell of each ice form, 
eight (ice VIII) and twelve (ice IX)
calculations were performed
where only one of the water molecules was included in turn in the QM region.
The cell's 
EFG results were then obtained by averaging over the unit cell's 8 or 12 water molecules. 
The static multipoles and polarizabilities of the environmental waters were calculated with the Dalton program package~\cite{dalton} at the CAM-B3LYP/loprop-aug-cc-pVTZ~\cite{gagliardi2004local} level.

The QM calculations were carried out with the conventional CASSCF and with the VQE-SCF method. The same active space was used for both methods, with six electrons in six orbitals. The CASSCF calculations were carried out with  PySCF~\cite{Sun2017,Sun2020}, and the VQE-SCF calculations were performed with the Aurora program package~\cite{aurora}, which uses PySCF as a backend. We utilized the CPPE~\cite{scheurer2019cppe} module for the PE calculations. The fermionic ADAPT-VQE~\cite{grimsley2019adaptive} ansatz was used to carry out all the quantum computing simulations in the active space, including only single and double operators in the operator pool, as implemented in the Aurora program package~\cite{aurora}.


The QM calculations were carried out using the Dunning basis sets cc-pV$X$Z, aug-cc-pV$X$Z, cc-pCV$X$Z, and aug-cc-pCV$X$Z,~\cite{dunning1989a, kendall1992a, Woon1995} with a cardinal number
$X$ = D, T and Q. 
A few representatives of the Pople basis sets have also been used, namely 6-311++G and 6-311++G**~\cite{krishnan1980a, clark1983a}. The wave function gradient norm was converged to $8\times 10^{-5}$, and the energy to $10^{-6}$. These convergence parameters were chosen arbitrarily based on a few sample calculations when we performed a hundred iterations and checked how much the energy and the gradient changed with each iteration. We used the L-BFGS-B optimizer for the quantum simulations. The PE-VQE-SCF calculations included the effect of shot noise during both the optimization and the measurement of the properties, using $10^5$ shots for all calculations. 
To calculate the NQIs from the EFG, we used the nuclear quadrupole moments $Q$ of $0.0028578\cdot10^{-28}$ for deuterium and $0.0256\cdot10^{-28}$ for $\mathrm{O}^{17}$~\cite{NEQM_table}.

To show the contribution of the PE environment to the final results, we also carried out calculations for single water molecules in a vacuum, without including any environment. Since the geometries are slightly different for the different water molecules in the unit cell, we carried out calculations for each one of them and took the average.



\section{Results}
\label{Sec:Results}

\subsection{Ice VIII}
In this section, we start by looking at the results obtained for the ice VIII system with the different basis sets.
Because of its $r^{-3}$ dependency, the EFG is a challenging property to calculate accurately, since it is hard to converge it to the basis set limit even when using larger basis sets~\cite{halkier1997}. In other works~\cite{halkier1997, Aidas2013}, the most accurate results have been obtained when using core‐valence basis sets such as cc-pCVXZ and aug-cc-pCVXZ.\\

\begin{table}[h!]
\centering 
\caption{Ice VIII. Deviations of the calculated NQI from the experimental results in kHz as a function of the basis set and with the inclusion of the environment. The experimental values used as reference are (7140 $\pm$ 100)~kHz for oxygen and 
(236.2 $\pm$ 0.3)~kHz for hydrogen~\cite{Edmonds1977}.
The calculated NQI values can be found in the Supplementary Information, Tables~\ref{SI-tab:ice8_NQI_6-311++G}--\ref{SI-tab:ice8_NQI_aug-cc-pCVTZ}.
}
\label{tab:ice8dev}
\small
\begin{tabular}{lcccc} \toprule 
& Classical & Simulator & Classical & Simulator\\
Basis & Oxygen & Oxygen & Hydrogen & Hydrogen\\[1.0ex]

\hline6-311++G & 1220 & 1216 & 58.51 & 58.51 \\ [1.0ex]
6-311++G** & 432.6 & 425.7 & 37.18 & 37.11 \\ [1.0ex]
aug-cc-pVDZ & $-$43.91 & - &34.11 & - \\ [1.0ex]
aug-cc-pVTZ & 292.9 & - & 34.83 & - \\ [1.0ex]
aug-cc-pVQZ & 373.2 & - & 19.89 & - \\ [1.0ex]
cc-pCVDZ & 1570 & 1666 & 40.72 & 40.70 \\ [1.0ex]
cc-pCVTZ & 874.2 & 901.0 & 35.91 & 35.66 \\ [1.0ex]
aug-cc-pCVDZ & 44.98 & 31.24 & 34.49 & 33.84 \\ [1.0ex]
aug-cc-pCVTZ & 183.6 & 191.8 & 34.57 & 34.44 \\ [1.0ex]

\bottomrule \end{tabular}\end{table} \normalsize

In our calculations, we find that diffuse basis functions play a crucial role in reproducing the experimental results. As shown in Table \ref{tab:ice8dev}, by adding diffuse functions to the cc-pCVDZ basis set, we get closer to the desired experimental values of the NQI by 
1525.02~kHz for oxygen and 
6.23~kHz for hydrogen in the classical case, and by 1634.76~kHz for oxygen and 6.23~kHz for hydrogen using a simulator. A similar trend is observed when the basis set changes from cc-pCVTZ to aug-cc-pCVTZ. We obtain the best results using either the aug-cc-pVDZ or aug-cc-pCVDZ basis sets, where the predicted results are in both cases within the experimental uncertainty for oxygen ($\pm 100$ kHz). When using very large basis sets, the quality of the results worsens. This could be due to several factors; for example, the smaller basis sets may benefit from fortuitous error cancellation, compensate for errors in the wave-function model, the employed geometries, and the environment description, among others. \\

\begin{figure}[ht!]
    \centering
    \subfloat{\includegraphics[width=8cm]{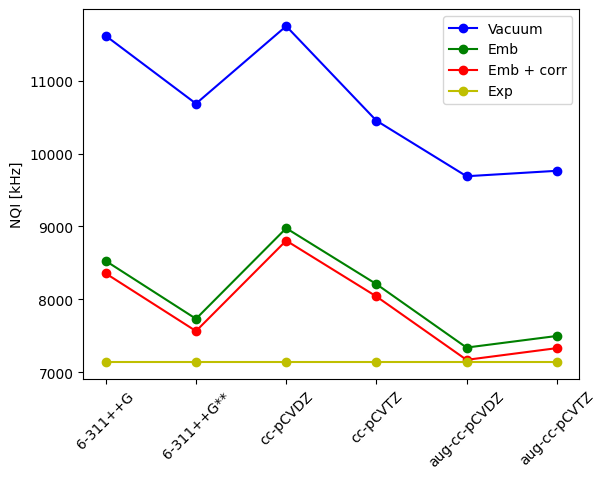}}
    \subfloat{\includegraphics[width=8cm]{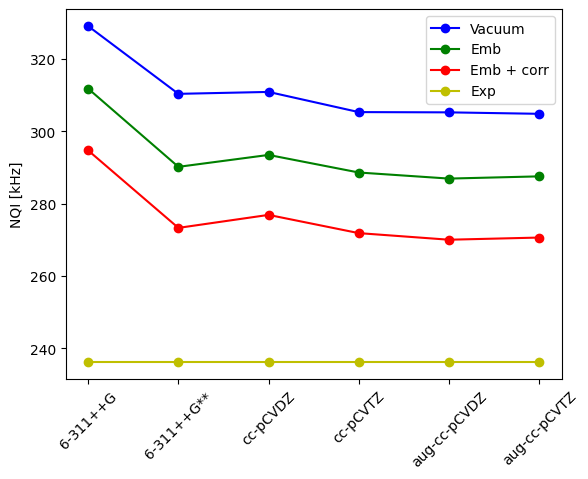}}
    \caption{Ice VIII.
NQI results calculated with the VQE-SCF method. Left: oxygen results; Right:  hydrogen results. Blue: calculations carried out in vacuum; Green: calculations carried out in a PE environment; Red: including the direct contribution of the PE environment to the EFG; 
Yellow: experimental results.
} \label{fig:ice8_QCC_column}
\end{figure}

As shown in Figure \ref{fig:ice8_QCC_column} (which is based on the results  in Tables
~\ref{SI-tab:ice8_NQI_6-311++G}--\ref{SI-tab:ice8_NQI_aug-cc-pCVTZ}), the inclusion of the environment considerably improves the results of our calculations toward the experimental values for both hydrogen and oxygen. The average overall improvement is $2779.70$ kHz for oxygen and $34.72$ kHz for hydrogen, which is $38.9\%$ and $14.7\%$ of the experimental result for oxygen and hydrogen, respectively. The inclusion of the environment without the direct contribution to the EFG adds $2610.69$ kHz for oxygen and $17.86$ kHz for hydrogen; the average direct contribution is $169.02$ kHz and $16.85$ kHz for oxygen and hydrogen, respectively, which is the $6.08\%$ and $48.5\%$ of the overall improvement. Here the direct contribution refers to the classically calculated contribution to the EFG arising directly from the presence of the solvent molecules. \\

The effect of the inclusion of the environment differs for different basis sets for oxygen, whereas for hydrogen it is between 16--20 kHz for every basis set, which corresponds to $6.7-8.4\%$ of the experimental value. 
For oxygen, the largest difference is found with the 6-311++G and 6-311++G** basis sets: $3086.9$ kHz and $2950.6$ kHz, that is $43.2\%$ and $41.3\%$ of the experimental value, respectively. The smallest differences are with the cc-pCVTZ, aug-cc-pCVDZ, and aug-cc-pCVTZ basis sets: 2241.2~kHz ($31.4\%$), 
2349.5~kHz ($32.9\%$), and 2264.9~kHz 
($31.7\%$), respectively. The direct contribution to the EFG tensor from the PE environment 
gives a smaller but not negligible improvement. It varies between 168--172 kHz for oxygen ($2.3$--$2.4\%$), and 16--17 kHz for hydrogen ($6.7$--$7.2\%$) for each basis set.\\

Next, we compare the NQI results obtained with the different methods and the inclusion of the PE environment.
Generally, the NQI results obtained with a simulator are close to the CASSCF results (within $50$ kHz). 
There is a larger difference upon the use of the cc-pCVDZ basis set due to two outliers, where there is a larger difference in energy and in NQC from the other 6 calculations. 
The calculated energies can be found in the 
Supplementary 
Information \cite{supp}, see Tables~\ref{SI-tab:ice8_energy_6-311++G}--\ref{SI-tab:ice8_energy_aug-cc-pCVTZ}.  Excluding these two results from the average, the obtained deviation from the experimental NQI is $1571$ kHz for oxygen and $40.37$ kHz for hydrogen, which lies very close to the classically obtained deviations, which is $1570$ kHz for oxygen and $40.70$ kHz for hydrogen. 
There are two further outliers with the cc-pCVTZ basis.
Excluding these two calculations from the average, we get $879.2$ kHz and $35.56$ kHz as a deviation from the experiment for oxygen and hydrogen, respectively. This brings the average for oxygen even closer to the CASSCF calculation ($874.2$ kHz). 
The experimental results could be reproduced with the aug-cc-pVDZ and the aug-cc-pCVDZ basis sets for oxygen.\\

Since we only include the most important excitations in our ADAPT-VQE simulation originating from a pool of single- and double excitation operators, we do not have exactly the same wavefunction as in the case of conventional CASSCF calculation, where we instead include every excitation in the active space. 
Furthermore, during the simulations, we also included shot noise. Therefore, in principle, we expect slightly worse results for the simulations. However, with some error cancellation, it is possible to obtain better results that lie closer to the experimental results than with the conventional method.\\

Not surprisingly, also in the case of the asymmetry parameters
(cf. Tables~\ref{SI-tab:ice8_asym_6-311++G}--\ref{SI-tab:ice8_asym_aug-cc-pCVTZ}), the inclusion of diffuse functions plays an important role. 
The addition of these functions improves the results for oxygen and hydrogen for both methods.\\[2mm]

The inclusion of an environment 
contributes to the parameters by $-0.108$ for oxygen and $0.027$ for hydrogen on average, which corresponds to $11.1\%$ and 
$26.5\%$ of the experimental value, respectively. This improvement is different for the different basis sets, and a trend similar to the one observed in the case of NQI contributions
can be seen. 
The largest improvements for oxygen occur using the 6-311++G ($-0.124$) and 6-311++G** ($-0.130$) basis sets, the smallest ones are found using the cc-pCVTZ ($-0.085$) and aug-cc-pCVTZ ($-0.097$) basis sets. For hydrogen, the trend is different, here the largest changes are for aug-cc-pCVDZ ($0.035$) and aug-cc-pCVTZ ($0.030$), and the smallest is for cc-pCVTZ ($0.021$).\\[2mm]

In general, the 
results obtained from the quantum simulations
are close to the CASSCF results, as 
shown in Table \ref{tab:ice8asym}. The parameters for oxygen lie outside the experimental uncertainty except for the 6-311++G** basis set. There is a $0.047$ and $0.041$ deviation from the experimental value 
with aug-cc-pCVDZ for the classical calculation and the quantum simulation, respectively. For hydrogen, only the 6-311++G and the aug-cc-pCVTZ results reproduce the experimental results; in the aug-cc-pCVDZ basis, the deviations from the experiment are $0.007$ and $0.006$,  for the CASSCF and the VQE-SCF, respectively.

\begin{table}[h!]
\centering 
\caption{Ice VIII. 
Asymmetry parameters $\eta_K$ calculated using different basis sets with the inclusion of the environment, and comparison with experimental values from Ref.~\cite{Edmonds1977}.}
\label{tab:ice8asym}
\small
\begin{tabular}{lcccc} \toprule 
& Classical & Simulator & Classical & Simulator\\
Basis & Oxygen & Oxygen & Hydrogen & Hydrogen\\[1.0ex]

\hline6-311++G &  0.926 &  0.927 &  0.103 &  0.103 \\ [1.0ex]
6-311++G** &  0.941 &  0.942 &  0.114 &  0.114 \\ [1.0ex]
cc-pCVDZ &  0.882 &  0.877 &  0.117 &  0.116 \\ [1.0ex]
cc-pCVTZ &  0.899 &  0.901 &  0.114 &  0.113 \\ [1.0ex]
aug-cc-pCVDZ &  0.923 & 0.930 &  0.109 &  0.108 \\ [1.0ex]
aug-cc-pCVTZ &  0.924 &  0.927 &  0.107 &  0.106 \\ [1.0ex]
\hline
Experimental & 0.97$\pm$0.03 &  0.97$\pm$0.03 & 0.102$\pm$0.005 & 0.102$\pm$0.005   \\ [1.0ex]

\bottomrule \end{tabular}\end{table} \normalsize

\subsection{Ice IX}

The results for the NQI deviations from the experiment 
of ice IX 
are collected in Table \ref{tab:ice9dev}. 
The full set of computed values can be found in 
Tables~\ref{SI-tab:ice9_NQI_6-311++G}--
\ref{SI-tab:ice9_NQI_aug-cc-pCVTZ}.

In the case of ice IX, both diffuse and core-valence correlated functions seem to play an important role in reproducing the experimental NQI results. We get the best results when using the aug-cc-pCVDZ and aug-cc-pCVTZ basis sets. The measurement uncertainty on the NQI of ice IX is $\pm 10$ kHz for oxygen and $\pm 3$ kHz for hydrogen. Even though our calculations could not reach this, the computed results lie even closer to the experimental results than in the case of ice VIII. 
When using the aug-cc-pVQZ basis set, we observe a reduction in the quality of our predictions compared to the experiment values. \\[2mm]

\begin{figure}[ht!]
    \centering
    \subfloat{\includegraphics[width=8cm]{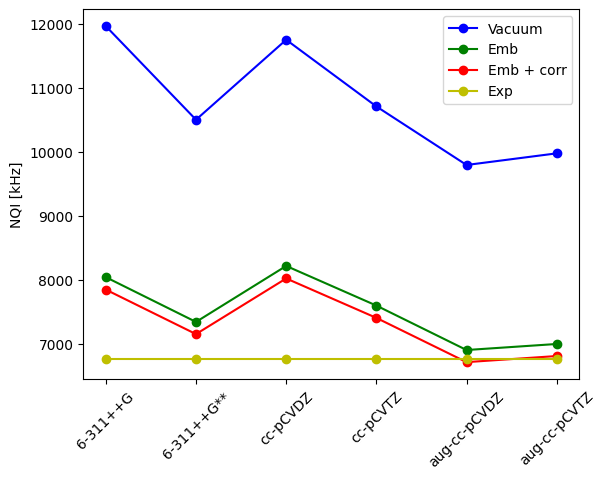}}
    \subfloat{\includegraphics[width=8cm]{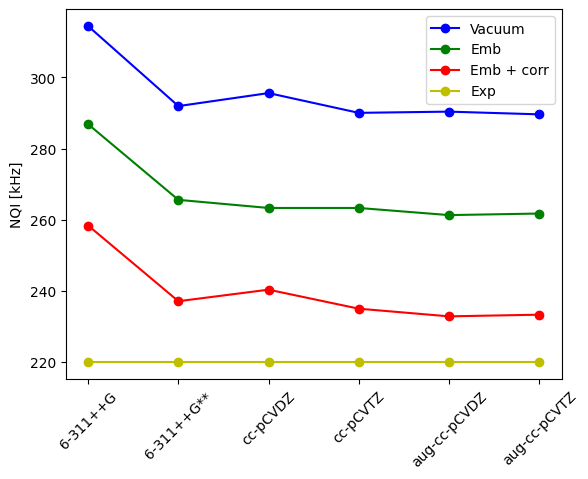}}
    \caption{Ice IX. NQI results obtained with the VQE-SCF 
    method and different basis sets. Left: results for oxygen, right: results for hydrogen. Blue: calculations carried out in vacuum; green: calculations carried out in a PE environment; red: including the direct contribution of the PE environment to the EFG; yellow: experimental results.}
    \label{fig:ice9_QCC_column}
\end{figure}

As can be seen in 
Figure~\ref{fig:ice9_QCC_column}, and as shown previously for ice VIII, the inclusion of the environment is crucial to reproduce the experimental data for the NQI. Here, the overall contribution of the environment is even larger than for ice VIII; $3456$ kHz for oxygen and $55.81$ kHz for hydrogen which correspond to $51.1\%$ and $25.4\%$ of the experimental result. 
The inclusion of the environment without the direct contribution improves the NQI by $3264$ kHz and $27.46$ kHz on average for oxygen and hydrogen, respectively, which is the $94.4\%$ and the $49.2\%$ of the overall contribution of the environment. The average direct contribution adds $191.5$ kHz for oxygen and $28.35$ kHz for hydrogen, which is $5.54\%$ and $50.8\%$ of the overall improvement. The indirect contributions from the environment vary for oxygen with the different basis sets, the largest change can be noticed with the 6-311++G ($3920$ kHz) and cc-pCVDZ ($3532$ kHz) basis sets. The smallest change can be observed with the aug-cc-pCVDZ ($2889$ kHz) and aug-cc-pCVTZ ($29.07$ kHz) basis sets, similar to the case of ice VIII. The overall contribution for hydrogen varies between 54--58 kHz, where the direct and indirect contribution is around 50--50\% of the overall environmental contribution for all basis sets.\\

In general, there is a larger difference between the CASSCF and ADAPT-VQE calculations in the obtained energies as compared to ice VIII, where the energy difference was within chemical accuracy for most of the calculations. These results can be found in the 
Supplementary Information~\cite{supp},
Tables~\ref{SI-tab:ice9_energy_6-311++G}--\ref{SI-tab:ice9_energy_aug-cc-pCVTZ}. 
The NQI results, on the other hand, lie close to the CASSCF results in almost all cases. We obtained the best results using the aug-cc-pCVDZ or aug-cc-pCVTZ basis sets. 

\begin{table}[ht!]
\centering 
\caption{Ice IX. Deviations of the calculated NQI from the experimental results in kHz as a function of the basis set and with the inclusion of the environment. The experimental values used as reference are 
(6766 $\pm$ 10)~kHz for oxygen and 
(220 $\pm$ 3) kHz for hydrogen~\cite{Edmonds1977}.
The calculated NQI values can be found in the Supplementary Information, Tables~\ref{SI-tab:ice9_NQI_6-311++G}--\ref{SI-tab:ice9_NQI_aug-cc-pCVTZ}.
}
\label{tab:ice9dev}
\small
\begin{tabular}{lcccc} \toprule 
& Classical & Simulator & Classical & Simulator\\
Basis & Oxygen & Oxygen & Hydrogen & Hydrogen\\[1.0ex]

\hline6-311++G & 1090 & 1088 & $-$38.34 
& $-$38.36 \\ [1.0ex]
6-311++G** & 419.4 & 393.7 & 
$-$17.29 & $-$17.11 \\ [1.0ex]
aug-cc-pVDZ & $-$125.3 & - & 
$-$13.24 & - \\ [1.0ex]
aug-cc-pVTZ & 120.4 & - & $-$13.64 & - \\ [1.0ex]
aug-cc-pVQZ & 212.9 & - & 0.691 & - \\ [1.0ex]
cc-pCVDZ & 1260 & 1263 & $-$20.57 & $-$20.36 \\ [1.0ex]
cc-pCVTZ & 664.1 & 646.8 & $-$15.09 
& $-$15.01 \\ [1.0ex]
aug-cc-pCVDZ & $-$34.91 & $-$44.00 & $-$13.60 & $-$12.86 \\ [1.0ex]
aug-cc-pCVTZ & 32.47 & 51.36 & $-$13.46 & $-$13.33 \\ [1.0ex]

\bottomrule \end{tabular}\end{table} \normalsize

The usage of diffuse functions brings us closer to the desired experimental asymmetry parameters for both methods similarly as we could see it at the ice VIII system as well.

The inclusion of the environment improved our results here as well. The average improvement for oxygen is $-$0.030, and for hydrogen, it is $0.013$ that is the $\%3.3$ and $10.8\%$. For oxygen, the largest improvement occurred for the cc-pCVDZ basis ($-$0.041), the smallest for the 
6-311++G** ($-$0.024) and the 6-311++G (0.027) basis sets. There is a different trend for hydrogen where the largest change occurs with the aug-cc-pCVDZ (0.020) basis, and the smallest with the cc-pCVTZ (0.008) basis set.

The asymmetry parameters for hydrogen, see Table~\ref{tab:ice9asym} as well as Tables~\ref{SI-tab:ice9_asym_6-311++G}--\ref{SI-tab:ice9_asym_aug-cc-pCVTZ}, are within the experimental accuracy for each basis set we used. For oxygen, it is quite the opposite. There is a $0.064$ and $0.071$ difference upon using the aug-cc-pCVTZ basis for conventional CASSCF and VQE-SCF, that is, $7.1\%$ and $7.9\%$ of the experimental result, respectively. For the aug-cc-pCVDZ basis set, this difference is $0.07$ for both methods. 
\\

\begin{table}[ht!]
\centering 
\caption{Ice IX. Asymmetry parameters $\eta_K$ calculated with the inclusion of the environment, using different basis sets, and comparison with experimental values from Ref.~\cite{Edmonds1977}.}
\label{tab:ice9asym}
\small
\begin{tabular}{lcccc} \toprule 
& Classical & Simulator & Classical & Simulator\\
Basis & Oxygen & Oxygen & Hydrogen & Hydrogen\\[1.0ex]

\hline6-311++G &  0.802 & 0.803 &  0.119 &  0.119 \\ [1.0ex]
6-311++G** &  0.822 & 0.824 &  0.136 &  0.136 \\ [1.0ex]
cc-pCVDZ & 0.800 & 0.800 &  0.137 &  0.136 \\ [1.0ex]
cc-pCVTZ & 0.806 & 0.812&  0.133 &  0.132 \\ [1.0ex]
aug-cc-pCVDZ & 0.819 & 0.825 &  0.129 &  0.129 \\ [1.0ex]
aug-cc-pCVTZ & 0.832 & 0.825 &  0.126 &  0.127 \\ [1.0ex]
\hline
Experimental & 0.896 $\pm$ 0.007 &  0.896 $\pm$ 0.007 & 0.12 $\pm$ 0.03 & 0.12 $\pm$ 0.03   \\ [1.0ex]

\bottomrule \end{tabular}\end{table} \normalsize

\FloatBarrier

\subsection{Computational cost estimation on noisy hardware}
Finally, we consider the cost and the noise-resilience of our calculations via CNOT gate count.\\

\begin{figure}[ht!]
    \centering
    \subfloat{\includegraphics[width=14cm]{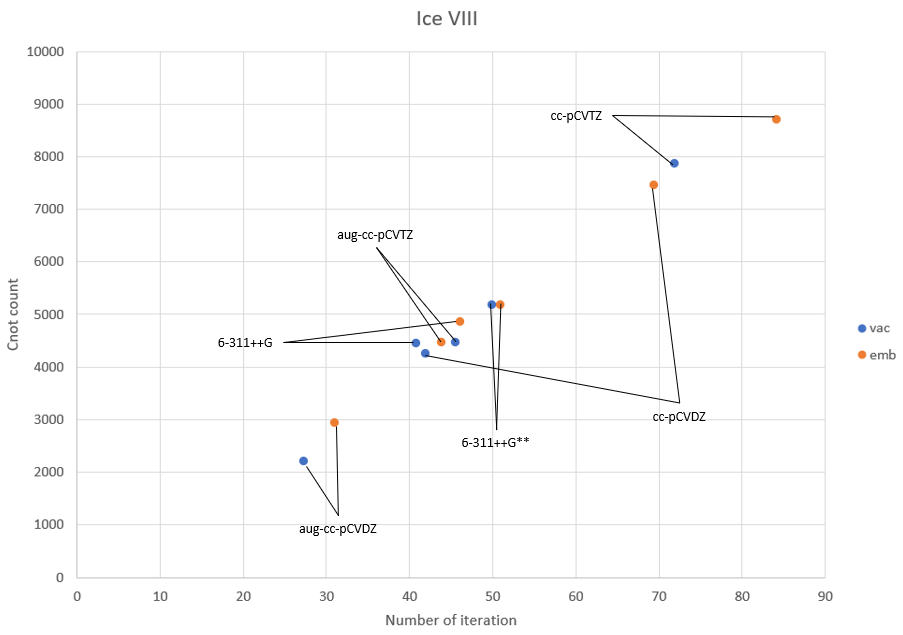}}
    \caption{The final CNOT count of the ansatz as a function of the iterations needed for convergence for the ice VIII system.}
    \label{fig:itercnot_ice8}
\end{figure}

\begin{figure}[htb!]
    \centering
    \subfloat{\includegraphics[width=14cm]{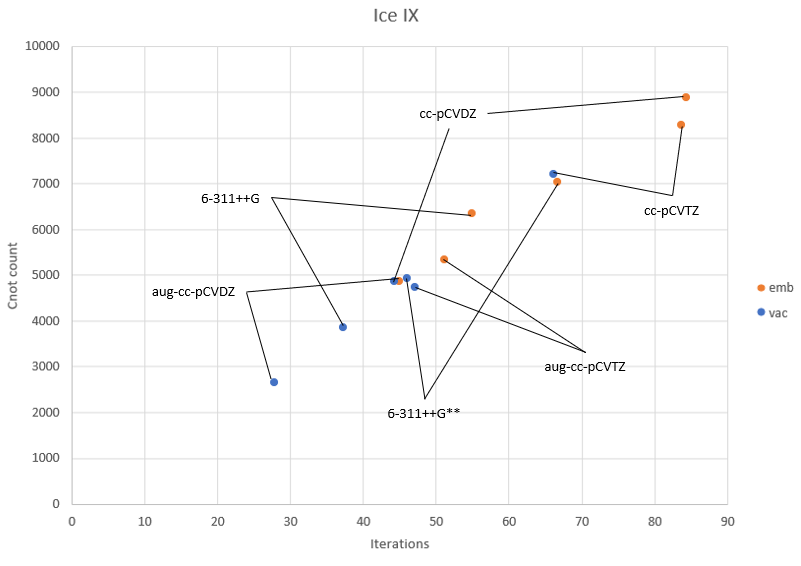}}
    \caption{The final CNOT count of the ansatz as a function of the iterations needed for convergence for the ice IX system.}
    \label{fig:itercnot_ice9}
\end{figure}

The average CNOT count as a function of the averaged iterations for the different basis sets can be seen 
in Figure \ref{fig:itercnot_ice8} and \ref{fig:itercnot_ice9}
(numerical values are in Tables
\ref{SI-tab:ice8_itervscnot}
and 
\ref{SI-tab:ice9_itervscnot}).
Within the ADAPT-VQE method, in every iteration, we add a new single or double excitation operator to the ansatz, so from the number of iterations, we know the number of operators that we have added. In both systems, the CNOT count seems to depend linearly on the number of iterations, regardless of the basis set or the addition of the environment. As it has been found before \cite{Kjellgren2023}, adding an environment to the calculation does not necessarily increase the CNOT count. However, it makes the wavefunction harder to converge. Therefore, it requires more terms in the ADAPT expansion and more CNOT gates. 
The addition of polarized or zeta functions to our basis set increased the number of iterations needed for convergence. However, the addition of diffuse functions did the opposite and lowered the computational cost upon using aug-cc-pCVDZ and aug-cc-pCVTZ basis sets. The previously mentioned most successful aug-cc-pCVDZ basis set needed the least terms to converge in both cases.\\

\pagebreak
\clearpage
\section{Conclusions}
In this work, NQI results have been calculated for ice VIII and IX with the classical CASSCF method and with a quantum simulator using ADAPT-VQE with a (6,6) active space in a polarizable embedding environment simulating 5832 or 8748 water
molecules in the environment aiming to reach the experimental accuracy. In the case of ice VIII using the aug-cc-pCVDZ basis set, the results of the calculations are within the experiment's uncertainty ($\pm 100$ kHz for O and $\pm 0.3$ kHz for H) for oxygen for both conventional CASSCF and VQE-SCF method. For ice IX, the aug-cc-pCVDZ and aug-cc-pCVTZ basis sets performed similarly well. In this case, the experimental uncertainty is much smaller ($\pm 10$ kHz for O and $\pm 3$kHz for H) then for ice VIII and therefore has not been reached, but the calculated results lie even closer to the experimental results than in the case of ice VIII.\\

In both systems, we observe a significant improvement in the results when a PE environment is applied. In the case of ice VIII with the inclusion of the environment, on average, we could recover an additional $2779$ kHz of the NQI (which is $38.9\%$ of the experimental result) for oxygen and $34.71$ kHz (which is $14.7\%$) for hydrogen. For ice IX on average, we recovered an additional $3456$ kHz for oxygen and 55.81 kHz for 
hydrogen, that is, 51.1\% and 25.4\% of the experimental NQI. This shows the importance of the inclusion of an environmental model.\\

The CNOT counts as the function of iterations have also been analyzed, which showed that neither the size of the basis sets nor the inclusion of an environment increased the CNOT count directly. Including an environment in the calculations makes the wavefunction harder to converge, and therefore, more operators need to be added from the operator pool, which, however, results in a higher CNOT count in general.

\printbibliography
\end{document}